\documentclass [11pt,a4paper] {article}

\usepackage[cp1252]{inputenc}
\usepackage{amssymb}
\usepackage{amsmath}
\usepackage{amsfonts,amssymb}
\usepackage[dvips]{graphicx}
\usepackage{bbm}
\usepackage{enumerate}
\usepackage{amsthm}
\usepackage{cancel}

\DeclareMathAlphabet{\mathpzc}{OT1}{pzc}{m}{it}

\def\SmallColSep{\setlength{\arraycolsep}{1pt}}

\newtheorem{conjecture}{Conjecture} 

\begin{document}

\title{Equal cost of computation for truth and falsity of experimental quantum propositions necessitates quantum parallel computing}

\author{Arkady Bolotin\footnote{$Email: arkadyv@bgu.ac.il$\vspace{5pt}} \\ \emph{Ben-Gurion University of the Negev, Beersheba (Israel)}}

\maketitle

\begin{abstract}\noindent Notwithstanding interest and excitement building around quantum computing in the last decades, a concise statement saying where this computing can truly help is still missing. As it is shown in the present paper, equal cost of computation for truth and falsity of experimental quantum propositions (required in order to infer a conclusion from a premise) cannot be achieved with classical computing. On the other hand, this equality might be realized with quantum parallel computing provided that the efficiency of such computing can be greater than 1.\\

\end{abstract}

\section{Introduction}  

\noindent Can a valuation, that is, an assignment of truth values, true and false, to experimental propositions pertaining to a quantum system, be regarded as a computational problem? Despite an apparent dissimilarity between valuation and computation, this question should be answered in affirmative.\\

\noindent Indeed, suppose that a quantum system is associated with a (separable) Hilbert space $\mathcal{H}$, whose unitary vectors correspond with possible pure states of the quantum system. Now recall that, in accordance with Birkhoff and von Neumann’s proposal \cite{Birkhoff}, the mathematical representative of an experimental proposition, denoted, for instance, by $P$, is a closed linear subspace of $\mathcal{H}$, say $\mathcal{P}$. So, if the system is in the pure state, which corresponds with the unit vector $|\Psi\rangle$ that either belongs or does not belong to the closed subspace $\mathcal{P}$, one can say that $|\Psi\rangle$ assigns a truth value of either true or false to the proposition $P$.\\

\noindent On the other hand, from elementary linear algebra (see, for example \cite{Anton}) it is known that $\mathcal{P}$ can be represented by the subspace spanned by linearly independent column vectors of the matrix $\mathbf{M}(\hat{P})$ that encodes – with respect to some arbitrary orthonormal basis – the projection operator $\hat{P}$ accordant to the proposition $P$. Consequently, the statement declaring that $|\Psi\rangle$ belongs to $\mathcal{P}$ is equivalent to the statement that the matrix $\mathbf{\Psi}$ encoding the unitary vector $|\Psi\rangle$ with respect to the chosen basis is within the said span. Otherwise stated, the statement $|\Psi\rangle\in\mathcal{P}$ is true together with the statement asserting the solvability of the system of linear equations $\mathbf{RX}\mkern-3mu=\mkern-2mu\mathbf{\Psi}$, where $\mathbf{R}$ is the matrix that comprises the linearly independent column vectors of $\mathbf{M}(\hat{P})$.\\

\noindent In this way, the assignment of truth values to experimental quantum propositions possesses the cost of computation which can be measured in the amount of time taken to check the solvability of systems of linear equations such as $\mathbf{RX}\mkern-3mu=\mkern-2mu\mathbf{\Psi}$.\\

\noindent The crucial question here is this: Does the cost of computation determined for the value of true differ from that determined for the value of false?\\

\noindent Where the response to this question to be positive, it would imply that it was not possible to infer a conclusion from a premise in accordance with a standard propositional calculus.\\

\noindent But then again, if the response to this question were to be negative, i.e., both truth values of any experimental quantum proposition were of equal cost of computation, it would entail the existence of based on quantum mechanics parallel computing whose efficiency – unlike that of classical parallel computing – could be greater than 1.\\

\noindent The present paper demonstrates the inference of those conclusions in detail.\\

\section{Preliminaries}  

\noindent Recall that any closed linear subspace of $\mathcal{H}$, for instance $\mathcal{P}$, is the range or the kernel of some projection operator $\hat{P}$ acting on $\mathcal{H}$ \cite{Kalmbach}, explicitly,\smallskip

\begin{equation} \label{CLS} 
   \mathcal{P}
   =
   \left\{
      \begingroup\SmallColSep
      \begin{array}{r l}
         \mathrm{ran}(\hat{P})
         ,
         &
         \mkern15mu
         \text{if}
         \mkern3mu
         \left\{
            |\varphi\rangle
            \in
            \mathcal{P}
            \textnormal{:}
            \mkern10mu
            \hat{P}
            |\varphi\rangle
            =
            |\varphi\rangle
            \mkern-2mu    
         \right\}
         \mkern-2mu
         \neq
         \mkern-2mu
         \varnothing
         \mkern6mu
         \text{is true}
         \\[10pt]
         \mathrm{ker}(\hat{P})
         ,
         &
         \mkern15mu
         \text{if}
         \mkern3mu
         \left\{
            |\varphi\rangle
            \in
            \mathcal{P}
            \textnormal{:}
            \mkern10mu
            \hat{P}
            |\varphi\rangle
            =
            0
            \mkern-2mu    
         \right\}
         \mkern-2mu
         \neq
         \mkern-2mu
         \varnothing
         \mkern6mu
         \text{is true}
      \end{array}
      \endgroup   
   \right.
   \;\;\;\;  .
\end{equation}
\smallskip

\noindent In view of that, the zero subspace $\{0\}$ and the identity subspace $\mathcal{H}$ are the closed linear subspaces of $\mathcal{H}$, namely, $\{0\}\mkern-2mu =\mkern-1mu\mathrm{ran}(\hat{0})\mkern-2mu=\mkern-1mu\mathrm{ker}(\hat{1})$ and $\mathcal{H}\mkern-2mu=\mkern-1mu\mathrm{ran}(\hat{1})\mkern-2mu=\mkern-1mu\mathrm{ker}(\hat{0})$, where $\hat{0}$ and $\hat{1}$ are the zero and identity operators, correspondingly.\\

\noindent As every unit vector $|\Psi\rangle\in\mathcal{H}$ can be decomposed uniquely as $|\Psi\rangle=|\psi\rangle+|\phi\rangle$ with $|\psi\rangle=\hat{P}|\Psi\rangle$ and $|\phi\rangle=|\Psi\rangle-\hat{P}|\Psi\rangle=(\hat{1}-\hat{P})|\Psi\rangle$, where $|\psi\rangle\in\mathrm{ran}(\hat{P})$ and $|\phi\rangle\in\mathrm{ker}(\hat{P})$, the subspaces $\mathrm{ran}(\hat{P})$ and $\mathrm{ker}(\hat{P})$ decompose the Hilbert space $\mathcal{H}$ into the direct sum:\smallskip

\begin{equation}  
   \mathcal{H}
   =
   \mathrm{ran}(\hat{P})
   \oplus
   \mathrm{ker}(\hat{P})
   \;\;\;\;  .
\end{equation}
\smallskip

\noindent One can infer from this fact that if the unit vector $|\Psi\rangle$ belongs to $\mathrm{ran}(\hat{P})$, then $|\Psi\rangle$ does not belong to $\mathrm{ker}(\hat{P})$ and vice versa.\\

\noindent In line with Birkhoff and von Neumann’s proposal \cite{Birkhoff}, let the mathematical representative of an experimental proposition $P$ be $\mathrm{ran}(\hat{P})$, the range of the projection operator $\hat{P}$ accordant to the proposition $P$. Likewise, let the mathematical representative of $\neg{P}$, the negation of the proposition $P$, be $\mathrm{ker}(\hat{P})$, the kernel of the projection operator $\hat{P}$.\\

\noindent Let us use the double-bracket notation ${[\mkern-3.3mu[\cdot]\mkern-3.3mu]}_v$ to express a truth value of an experimental atomic proposition or a propositional formula constructed from experimental atomic propositions.\\

\noindent The truth value ${[\mkern-3.3mu[P]\mkern-3.3mu]}_v$ of the experimental atomic proposition $P$ in the pure state described by the unit vector $|\Psi\rangle$ of $\mathcal{H}$ can be determined by the agency of the relation ``is an element of'' between the vector $|\Psi\rangle$ and the closed linear subspaces $\mathrm{ran}(\hat{P})$ and $\mathrm{ker}(\hat{P})$.\\

\noindent Let the map\smallskip

\begin{equation} \label{MAP} 
   v
   \mkern-3.3mu
   :
   \mkern2mu
   \mathbb{B}_{2}
   \times
   \mathbb{B}_{2}
   \to
   \mathbb{B}_{2}
   \;\;\;\;  ,
\end{equation}
\smallskip

\noindent where $\mathbb{B}_{2}$ is the set of two truth values, \emph{true} and \emph{false} (or 1 and 0), be the function\smallskip

\begin{equation}  
   {[\mkern-3.3mu[P]\mkern-3.3mu]}_v
   =
   v
   \left(
      |\Psi\rangle
      \mkern-2.5mu
      \in
      \mkern-2mu
      \mathrm{ran}(\hat{P})
      ,
      |\Psi\rangle
      \mkern-2.5mu
      \in
      \mkern-2mu
      \mathrm{ker}(\hat{P})
   \right)
   \;\;\;\;   
\end{equation}
\smallskip

\noindent that takes the mathematical statements $|\Psi\rangle\mkern-2.5mu\in\mkern-2mu\mathrm{ran}(\hat{P})$ and $|\Psi\rangle\mkern-2.5mu\in\mkern-2mu\mathrm{ker}(\hat{P})$, which have been described as true or false, to elements of $\mathbb{B}_{2}$. Specifically, $v$ returns the truth value of true if the statement $|\Psi\rangle\mkern-2.5mu\in\mkern-2mu\mathrm{ran}(\hat{P})$ is true (and hence $|\Psi\rangle\mkern-2.5mu\in\mkern-2mu\mathrm{ker}(\hat{P})$ is false), and $v$ returns the truth value of false if the statement $|\Psi\rangle\mkern-2.5mu\in\mkern-2mu\mathrm{ker}(\hat{P})$ is true (and so $|\Psi\rangle\mkern-2.5mu\in\mkern-2mu\mathrm{ran}(\hat{P})$ is false). In symbols,\smallskip

\begin{equation}  
   v
   \left(
      |\Psi\rangle
      \mkern-2.5mu
      \in
      \mkern-2mu
      \mathrm{ran}(\hat{P})
      ,
      |\Psi\rangle
      \mkern-2.5mu
      \in
      \mkern-2mu
      \mathrm{ker}(\hat{P})
   \right)
   =
   \left\{
      \begingroup\SmallColSep
      \begin{array}{r l}
         1
         ,
         &
         \mkern15mu
         \text{if}
         \mkern4mu
         |\Psi\rangle
         \mkern-2.5mu
         \in
         \mkern-2mu
         \mathrm{ran}(\hat{P})
         \mkern6mu
         \text{is true}
         \\[10pt]
         0
         ,
         &
         \mkern15mu
         \text{if}
         \mkern4mu
         |\Psi\rangle
         \mkern-2.5mu
         \in
         \mkern-2mu
         \mathrm{ker}(\hat{P})
         \mkern6mu
         \text{is true}
      \end{array}
      \endgroup   
   \right.
   \;\;\;\;  .
\end{equation}
\smallskip

\noindent For example, consider a spin-$^{1}\mkern-5mu/\mkern-3mu_{2}$ system, i.e., a qubit. In the Hilbert space $\mathbb{C}^{{2}\times{2}}$ of ${2}\times{2}$ complex matrices, the range and the kernel of the projection operator $\hat{Y}_{+^{1}\mkern-5mu/\mkern-3mu_{2}}$ accordant to the experimental atomic proposition ``The spin of the qubit along the $y$-axis is $+^{1}\mkern-5mu/\mkern-3mu_{2}\hbar\mkern3mu$”, denoted by $Y_{+^{1}\mkern-5mu/\mkern-3mu_{2}}$, are the closed linear subspaces\smallskip

\begin{equation}  
   \mathbb{Y}_{+^{1}\mkern-5mu/\mkern-3mu_{2}}
   =
   \left\{
      a
      \in
      \mathbb{C}
      \mkern-2mu
      :
      \mkern5mu
      \left[
         \begingroup\SmallColSep
         \begin{array}{r}
            a \\
            ia
         \end{array}
         \endgroup
      \right]
   \right\}
   \;\;\;\;  ,
\end{equation}
\\[-30pt]

\begin{equation}  
   \mathbb{Y}_{-^{1}\mkern-5mu/\mkern-3mu_{2}}
   =
   \left\{
      b
      \in
      \mathbb{C}
      \mkern-2mu
      :
      \mkern5mu
      \left[
         \begingroup\SmallColSep
         \begin{array}{r}
            ib \\
            b
         \end{array}
         \endgroup
      \right]
   \right\}
   \;\;\;\;  ,
\end{equation}
\smallskip

\noindent in that order. So, if the qubit is the pure state described by the unit vector of $\mathbb{C}^{{2}\times{2}}$\smallskip

\begin{equation}  
   \mathbf{\Psi}_{\mkern-2mu{1}}
   =
   \frac{1}{\sqrt{2}}
   \left[
      \begingroup\SmallColSep
      \begin{array}{r}
         1 \\
         i
      \end{array}
      \endgroup
   \right]
   \;\;\;\;  ,
\end{equation}
\smallskip

\noindent then the statement $\mathbf{\Psi}_{\mkern-2mu{1}}\mkern-2.5mu\in\mkern-2mu\mathbb{Y}_{+^{1}\mkern-5mu/\mkern-3mu_{2}}$ will be true (and the statement $\mathbf{\Psi}_{\mkern-2mu{1}}\mkern-2.5mu\in\mkern-2mu\mathbb{Y}_{-^{1}\mkern-5mu/\mkern-3mu_{2}}$ will be false); consequently, ${[\mkern-3.3mu[Y_{+^{1}\mkern-5mu/\mkern-3mu_{2}}]\mkern-3.3mu]}_v=1$ in the state $\mathbf{\Psi}_{\mkern-2mu{1}}$. Similarly, if the qubit is in the state described by the unit vector $\mathbf{\Psi}_{\mkern-2mu{2}}$ of $\mathbb{C}^{{2}\times{2}}$, namely,\smallskip

\begin{equation}  
   \mathbf{\Psi}_{\mkern-2mu{2}}
   =
   \frac{1}{\sqrt{2}}
   \left[
      \begingroup\SmallColSep
      \begin{array}{r}
         1 \\
         -i
      \end{array}
      \endgroup
   \right]
   \;\;\;\;  ,
\end{equation}
\smallskip

\noindent the statement $\mathbf{\Psi}_{\mkern-2mu{2}}\mkern-2.5mu\in\mkern-2mu\mathbb{Y}_{-^{1}\mkern-5mu/\mkern-3mu_{2}}$ is true (but $\mathbf{\Psi}_{\mkern-2mu{2}}\mkern-2.5mu\in\mkern-2mu\mathbb{Y}_{+^{1}\mkern-5mu/\mkern-3mu_{2}}$ comes to be false); hence, ${[\mkern-3.3mu[Y_{+^{1}\mkern-5mu/\mkern-3mu_{2}}]\mkern-3.3mu]}_v=0$ in $\mathbf{\Psi}_{\mkern-2mu{2}}$.\\

\noindent Yet, the third option exists, namely, both $|\Psi\rangle\mkern-2.5mu\in\mkern-2mu\mathrm{ran}(\hat{P})$ and $|\Psi\rangle\mkern-2.5mu\in\mkern-2mu\mathrm{ker}(\hat{P})$ are false. For example, the unit vector of $\mathbb{C}^{{2}\times{2}}$\smallskip

\begin{equation}  
   \mathbf{\Psi}_{\mkern-2mu{3}}
   =
   \left[
      \begingroup\SmallColSep
      \begin{array}{r}
         1 \\
         0
      \end{array}
      \endgroup
   \right]
   \;\;\;\;   
\end{equation}
\smallskip

\noindent belongs to neither $\mathbb{Y}_{+^{1}\mkern-5mu/\mkern-3mu_{2}}$ nor $\mathbb{Y}_{-^{1}\mkern-5mu/\mkern-3mu_{2}}$. This constitutes a problem since three options are incompatible with the map (\ref{MAP}) admitting only two truth values.\\

\noindent One workaround is to demand that under the function $v$, the truth of the mathematical statement $|\Psi\rangle\mkern-2.5mu\notin\mkern-2mu\mathrm{ran}(\hat{P})\sqcap|\Psi\rangle\mkern-2.5mu\notin\mkern-2mu\mathrm{ker}(\hat{P})$ (where the symbol $\sqcap$ stands for logical conjunction) has the same image as either the truth of the statement $|\Psi\rangle\mkern-2.5mu\in\mkern-2mu\mathrm{ran}(\hat{P})$ or the truth of the statement $|\Psi\rangle\mkern-2.5mu\in\mkern-2mu\mathrm{ker}(\hat{P})$ does (more details on this proposal will follow in the last section of the paper).\\

\noindent The alternative, which will be considered in this paper, is to deem the third option as the case where the function $v$ is undetermined and so the experimental quantum proposition $P$ has no truth value at all. In symbols,\smallskip

\begin{equation} \label{SV} 
   v
   \left(
      |\Psi\rangle
      \mkern-2.5mu
      \in
      \mkern-2mu
      \mathrm{ran}(\hat{P})
      ,
      |\Psi\rangle
      \mkern-2.5mu
      \in
      \mkern-2mu
      \mathrm{ker}(\hat{P})
   \right)
   =
   \left\{
      \begingroup\SmallColSep
      \begin{array}{r l}
         1
         ,
         &
         \mkern15mu
         \text{if}
         \mkern4mu
         |\Psi\rangle
         \mkern-2.5mu
         \in
         \mkern-2mu
         \mathrm{ran}(\hat{P})
         \mkern6mu
         \text{is true}
         \\[5pt]
         0
         ,
         &
         \mkern15mu
         \text{if}
         \mkern4mu
         |\Psi\rangle
         \mkern-2.5mu
         \in
         \mkern-2mu
         \mathrm{ker}(\hat{P})
         \mkern6mu
         \text{is true}
         \\[5pt]
         0/0
         ,
         &
         \mkern15mu
         \text{if}
         \mkern4mu
         |\Psi\rangle
         \mkern-2.5mu
         \notin
         \mkern-2mu
         \mathrm{ran}(\hat{P})
         \sqcap
         |\Psi\rangle
         \mkern-2.5mu
         \notin
         \mkern-2mu
         \mathrm{ker}(\hat{P})
         \mkern6mu
         \text{is true}
      \end{array}
      \endgroup   
   \right.
   \;\;\;\;  ,
\end{equation}
\smallskip

\noindent where $0/0$ symbolizes a truth-value gap.\\

\noindent E.g., as the statement $\mathbf{\Psi}_{\mkern-2mu{3}}\mkern-2.5mu\notin\mkern-2mu\mathbb{Y}_{+^{1}\mkern-5mu/\mkern-3mu_{2}}\sqcap\mathbf{\Psi}_{\mkern-2mu{3}}\mkern-2.5mu\notin\mkern-2mu\mathbb{Y}_{-^{1}\mkern-5mu/\mkern-3mu_{2}}$ is true, the proposition $Y_{+^{1}\mkern-5mu/\mkern-3mu_{2}}$ has no truth value in the state $\mathbf{\Psi}_{\mkern-2mu{3}}$.\\

\section{Valuation means computation}  

\noindent Consider a separable Hilbert space $\mathcal{H}$ of finite dimension $n$. Let the projection operator $\hat{P}$ acting on $\mathcal{H}$ be encoded by the complex ${n}\times{n}$ matrix $\mathbf{M}(\hat{P})$\smallskip

\begin{equation}  
   \mathbf{M}(\hat{P})
   =
   \mkern-2mu
   \left[
      \begingroup\SmallColSep
      \begin{array}{c c c}
         P_{11}
         &
         \cdots
         &
         P_{1n}
         \\
         \vdots
         &
         \ddots
         &
         \vdots
         \\
         P_{n1}
         &
         \cdots
         &
         P_{nn}
      \end{array}
      \endgroup
   \right]
   =
   \left(
      P_{ij}
   \right)_{i=1.\mkern2mu{j=1}}^{n.n}
   \in
   \mathbb{C}^{{n}\times{n}}
   \;\;\;\;   
\end{equation}
\smallskip

\noindent having the entries $P_{ij}$\smallskip

\begin{equation}  
   P_{ij}
   =
   \langle{e_i}|\hat{P}|{e_j}\rangle
   \;\;\;\;  ,
\end{equation}
\smallskip

\noindent where $|{e_j}\rangle$ are vectors of an arbitrary orthonormal basis $\{|{e_j}\rangle\}$, $\langle{e_i}|{e_j}\rangle=\delta_{ij}$, for $\mathcal{H}$. The range of the matrix $\mathbf{M}(\hat{P})$ is the same as the span of the column vectors $\mathbf{M}_j$ of this matrix, i.e.,\smallskip

\begin{equation}  
   \mathrm{ran}(\mathbf{M}(\hat{P}))
   =
   \mathrm{Span}
   \left(
      \mathbf{M}_{1}
      ,
      \dots
      ,
      \mathbf{M}_{n}
   \right)
   \;\;\;\;  ,
\end{equation}
\smallskip

\noindent where either $(\mathbf{M}_{1},\dots,\mathbf{M}_{n})$ is a basis for $\mathrm{ran}(\mathbf{M}(\hat{P}))$ or some $\mathbf{M}_{j}$ can be removed to obtain a basis for $\mathrm{ran}(\mathbf{M}(\hat{P}))$. Explicitly,\smallskip

\begin{equation}  
   \mathrm{ran}(\mathbf{M}(\hat{P}))
   =
   \left\{
      c_1
      ,
      \dots
      ,
      c_n
      \in
      \mathbb{C}
      :
      \mkern15mu
      c_1
      \mkern-4mu
      \left[
         \begingroup\SmallColSep
         \begin{array}{c}
            P_{11}  \\
            \vdots  \\
            P_{n1} 
         \end{array}
         \endgroup
      \right]
      +
      \cdots
      +
      c_n
      \mkern-4mu
      \left[
         \begingroup\SmallColSep
         \begin{array}{c}
            P_{1n}  \\
            \vdots  \\
            P_{nn} 
         \end{array}
         \endgroup
      \right]
   \right\}
   \;\;\;\;  .
\end{equation}
\smallskip

\noindent One can make here the following observation: The truth of the statement $|\Psi\rangle\mkern-2.5mu\in\mkern-2mu\mathrm{ran}(\hat{P})$ means the solvability of the system of linear equations\smallskip

\begin{equation}  
   \mathbf{RX}
   =
   \mathbf{\Psi}
    \;\;\;\;  ,
\end{equation}
\smallskip

\noindent in which $\mathbf{\Psi}$ is the column vector that contains the components of the unit vector $|\Psi\rangle$ with respect to the chosen basis $\{|{e_j}\rangle\}$, i.e.,\smallskip

\begin{equation}  
   \mathbf{\Psi}
   =
   \mathbf{M}(|\Psi\rangle)
   =
   \left(
      \psi_i
   \right)_{i=1}^{n}
   \in
   \mathbb{C}^{{n}\times{1}}
   \;\;\;\;  ,
\end{equation}
\smallskip

\noindent where\smallskip

\begin{equation}  
   \psi_{i}
   =
   \langle{e_i}|\Psi\rangle
   \;\;\;\;  ,
\end{equation}
\smallskip

\noindent while $\mathbf{X}\mkern-2mu\in\mkern-2mu\mathbb{C}^{{m}\times{1}}$ is the column vector with $m\le{n}$ entries $x_1,\dots,x_m$ which are put in the place of weights $c_1,\dots,c_m$ for the linearly independent column vectors $\mathbf{M}_{1},\dots,\mathbf{M}_{m}$ of the matrix $\mathbf{M}(\hat{P})$, so that\smallskip

\begin{equation}  
   \mathbf{RX}
   \equiv
   \left[
      \begingroup\SmallColSep
      \begin{array}{c}
         P_{11}  \\
         \vdots  \\
         P_{n1} 
      \end{array}
      \endgroup
   \right]
   \mkern-4mu
   x_1
   +
   \cdots
   +
   \left[
      \begingroup\SmallColSep
      \begin{array}{c}
         P_{1m}  \\
         \vdots  \\
         P_{nm} 
      \end{array}
      \endgroup
   \right]
   \mkern-4mu
   x_m
    \;\;\;\;  .
\end{equation}
\smallskip

\noindent Denoting $(\delta_{ij})_{i=1}^{n}\mkern-4mu\in\mkern-4mu\mathbb{C}^{{n}\times{1}}$ by $\mathbf{I}_{j}$, the kernel of $\mathbf{M}(\hat{P})$ can be presented as the span of the column vectors $\mathbf{I}_j-\mathbf{M}_j$ of the matrix $\mathbf{I}-\mathbf{M}(\hat{P})$, namely,\smallskip

\begin{equation}  
   \mathrm{ker}(\mathbf{M}(\hat{P}))
   =
   \mathrm{Span}
   \left(
      \mathbf{I}_{1}
      -
      \mathbf{M}_{1}
      ,
      \dots
      ,
      \mathbf{I}_{n}
      -
      \mathbf{M}_{n}
   \right)
   \;\;\;\;  ,
\end{equation}
\smallskip

\noindent where either $(\mathbf{I}_{1}-\mathbf{M}_{1},\dots,\mathbf{I}_{n}-\mathbf{M}_{n})$ is a basis for $\mathrm{ker}(\mathbf{M}(\hat{P}))$ or some $\mathbf{I}_{j}-\mathbf{M}_{j}$ can be removed to obtain a basis for $\mathrm{ker}(\mathbf{M}(\hat{P}))$; explicitly,\smallskip

\begin{equation}  
   \mathrm{ker}(\mathbf{M}(\hat{P}))
   =
   \left\{
      c_j
      \mkern-2mu
      \in
      \mkern-2mu
      \mathbb{C}
      :
      \mkern10mu
      c_1
      \mkern-4mu
      \left[
         \begingroup\SmallColSep
         \begin{array}{r}
            1-P_{11}  \\
            \vdots     \\
            -P_{n1} 
         \end{array}
         \endgroup
      \right]
      \mkern-4mu
      +
      \cdots
      +
      c_j
      \mkern-4mu
      \left[
         \begingroup\SmallColSep
         \begin{array}{r}
            \vdots                 \\
            \delta_{ij}-P_{ij}  \\
            \vdots 
         \end{array}
         \endgroup
      \right]
      \mkern-4mu
      +
      \cdots
      +
      c_n
      \mkern-4mu
      \left[
         \begingroup\SmallColSep
         \begin{array}{r}
            -P_{1n}  \\
            \vdots   \\
            1-P_{nn} 
         \end{array}
         \endgroup
      \right]
   \right\}
   \;\;\;\;  .
\end{equation}
\smallskip

\noindent In consequence, to decide whether $|\Psi\rangle$ belongs to $\mathrm{ker}(\hat{P})$ means to answer the question whether the following system of linear equations has at least one solution:\smallskip

\begin{equation}  
   \mathbf{KX}
   =
   \mathbf{\Psi}
    \;\;\;\;  ,
\end{equation}
\smallskip

\noindent where $\mathbf{X}\in\mkern-2mu\mathbb{C}^{{k}\times{1}}$ is the column vector with $k\le{n}$ entries $x_1,\dots,x_k$ which substitute weights $c_1,\dots,c_k$ for the linearly independent column vectors of the matrix $\mathbf{I}-\mathbf{M}(\hat{P})$ so that\smallskip

\begin{equation}  
   \mathbf{KX}
   \equiv
   \left[
      \begingroup\SmallColSep
      \begin{array}{r}
         1-P_{11}  \\
         \vdots     \\
         -P_{n1} 
      \end{array}
      \endgroup
   \right]
   \mkern-4mu
   x_1
   +
   \cdots
   +
   \left[
      \begingroup\SmallColSep
      \begin{array}{r}
         \vdots                 \\
         \delta_{ij}-P_{ij}  \\
         \vdots 
      \end{array}
      \endgroup
   \right]
   \mkern-4mu
   x_j
   +
   \cdots
   +
   \left[
      \begingroup\SmallColSep
      \begin{array}{r}
         -P_{1n}  \\
         \vdots   \\
         1-P_{nk} 
      \end{array}
      \endgroup
   \right]
   \mkern-4mu
   x_k
    \;\;\;\;  .
\end{equation}
\smallskip

\noindent \noindent Suppose that the unit vector $|\Psi\rangle$ lies in the range of the projection operator $\hat{P}$ with the result that $\hat{P}|\Psi\rangle=|\Psi\rangle$. This equation corresponds to the matrix equation\smallskip

\begin{equation}  
   \mathbf{M}(\hat{P})
   \mathbf{\Psi}
   =
   \mathbf{\Psi}
   \;\;\;\;  ,
\end{equation}
\smallskip

\noindent which indicates that $(\mathbf{M}(\hat{P}))^2=\mathbf{M}(\hat{P})$. Similarly, the bra equation $\langle\Psi|=\langle\Psi|\hat{P}$ can be written as\smallskip

\begin{equation}  
   \mathbf{\Psi}^{\dagger}
   =
   \mathbf{\Psi}^{\dagger}\mathbf{M}(\hat{P})
   \;\;\;\;  ,
\end{equation}
\smallskip

\noindent where the row vector $\mathbf{\Psi}^{\dagger}\mkern-2mu\in\mkern-2mu\mathbb{C}^{{1}\times{n}}$ having the $i^{\text{th}}$ entry $\langle\Psi|e_i\rangle$ is the adjoint matrix of the column vector $\mathbf{\Psi}$. From here, it follows that\smallskip

\begin{equation}  
   \mathbf{M}(\hat{P})
   \mathbf{\Psi}
   \mathbf{\Psi}^{\dagger}
   =
   \mathbf{\Psi}
   \mathbf{\Psi}^{\dagger}
   \mathbf{M}(\hat{P})
   =
   \mathbf{\Psi}
   \mathbf{\Psi}^{\dagger}
   \;\;\;\;  ,
\end{equation}
\smallskip

\noindent which can be if\smallskip

\begin{equation}  
   \mathbf{M}(\hat{P})
   =
   \mathbf{\Psi}
   \mathbf{\Psi}^{\dagger}
   \;\;\;\;  .
\end{equation}
\smallskip

\noindent As $\mathbf{\Psi}$ is a $n\times1$ matrix and $\mathbf{\Psi}^{\dagger}$ is a $1\times{n}$ matrix, the above factorization means (see, for example \cite{Mirsky}) that the rank of the matrix $\mathbf{M}(\hat{P})$ is $\mathrm{Rank}(\mathbf{M}(\hat{P}))=1$. This implies that the number of linearly independent column vectors of the matrix $\mathbf{M}(\hat{P})$ is 1, i.e.,\smallskip

\begin{equation}  
   \mathrm{Rank}(\mathbf{M}(\hat{P}))
   =
   \dim
   \left(
      \mathrm{Span}
      \left(
         \mathbf{M}_{1}
         ,
         \dots
         ,
         \mathbf{M}_{n}
      \right)
   \right)
   =
   m
   =
   1
   \;\;\;\;  .
\end{equation}
\smallskip

\noindent By the rank-nullity theorem \cite{Friedberg}, $\mathrm{Nullity}(\mathbf{M}(\hat{P}))=n-\mathrm{Rank}(\mathbf{M}(\hat{P}))$; so, for any $n\ge{2}$, the number of linearly independent column vectors of the matrix $\mathbf{I}-\mathbf{M}(\hat{P})$ is\smallskip

\begin{equation}  
   \mathrm{Nullity}(\mathbf{M}(\hat{P}))
   =
   \dim
   \left(
      \mathrm{Span}
      \left(
         \mathbf{I}_{1}
         -
         \mathbf{M}_{1}
         ,
         \dots
         ,
         \mathbf{I}_{n}
         -
         \mathbf{M}_{n}
      \right)
   \right)
   =
   k
   =
   n
   -
   1
   \;\;\;\;  .
\end{equation}
\smallskip

\noindent Stipulating that $U_{\mathbf{R}}$ is the solution set for the linear system $\mathbf{RX}\mkern-3mu=\mkern-2mu\mathbf{\Psi}$\smallskip

\begin{equation}  
   U_{\mathbf{R}}
   =
   \left\{
      \mathbf{X}
      \mkern-2mu
      \in
      \mkern-2mu
      \mathbb{C}
      \mkern-3mu
      :
      \mkern7.5mu
      \mathbf{RX}
      =
      \mathbf{\Psi}
   \right\}
   \;\;\;\;  ,
\end{equation}
\smallskip

\noindent and $U_{\mathbf{K}}$ is the solution set for the linear system $\mathbf{KX}\mkern-3mu=\mkern-2mu\mathbf{\Psi}$\smallskip

\begin{equation}  
   U_{\mathbf{K}}
   =
   \left\{
      \mathbf{X}
      \mkern-2mu
      \in
      \mkern-2mu
      \mathbb{C}^{{(n-1)}\times{1}}
      \mkern-3mu
      :
      \mkern7.5mu
      \mathbf{KX}
      =
      \mathbf{\Psi}
   \right\}
   \;\;\;\;  ,
\end{equation}
\smallskip

\noindent one can present the equivalence of the mathematical statements:\smallskip

\begin{equation}  
   \begingroup\SmallColSep
   \begin{array}{r l}
      |\Psi\rangle
      \mkern-2.5mu
      \in
      \mkern-2mu
      \mathrm{ran}(\hat{P})
      &
      \iff
      U_{\mathbf{R}}
      \neq
      \varnothing
      \\[5pt]
      |\Psi\rangle
      \mkern-2.5mu
      \in
      \mkern-2mu
      \mathrm{ker}(\hat{P})
      &
      \iff
      U_{\mathbf{K}}
      \neq
      \varnothing
      \\[5pt]
      \mkern15mu
      |\Psi\rangle
      \mkern-2.5mu
      \notin
      \mkern-2mu
      \mathrm{ran}(\hat{P})
      \sqcap
      |\Psi\rangle
      \mkern-2.5mu
      \notin
      \mkern-2mu
      \mathrm{ker}(\hat{P})
      &
      \iff
      U_{\mathbf{R}}
      =
      \varnothing
      \sqcap
      U_{\mathbf{K}}
      =
      \varnothing
   \end{array}
   \endgroup   
   \;\;\;\;  ,
\end{equation}
\smallskip

\noindent where the symbol $\iff$ stands for \emph{the logical biconditional} (which is true when its antecedent and consequent are either true or false at the same time).\\

\noindent Accordingly, the valuation defined in (\ref{SV}) can be rewritten in the form\smallskip

\begin{equation} \label{SV1} 
   {[\mkern-3.3mu[P]\mkern-3.3mu]}_v
   =
   v
   \Big(
      U_{\mathbf{R}}
      \neq
      \varnothing
      ,
      U_{\mathbf{K}}
      \neq
      \varnothing
   \Big)
   =
   \left\{
      \begingroup\SmallColSep
      \begin{array}{r l}
         1
         ,
         &
         \mkern15mu
         \text{if}
         \mkern6mu
         U_{\mathbf{R}}
         \neq
         \varnothing
         \mkern6mu
         \text{is true}
         \\[5pt]
         0
         ,
         &
         \mkern15mu
         \text{if}
         \mkern6mu
         U_{\mathbf{K}}
         \neq
         \varnothing
         \mkern6mu
         \text{is true}
         \\[5pt]
         0/0
         ,
         &
         \mkern15mu
         \text{if}
         \mkern6mu
         U_{\mathbf{R}}
         =
         \varnothing
         \sqcap
         U_{\mathbf{K}}
         =
         \varnothing
         \mkern6mu
         \text{is true}
      \end{array}
      \endgroup
   \right.
   \;\;\;\;  .
\end{equation}
\smallskip

\noindent As an illustration, consider a spin-$^{5}\mkern-5mu/\mkern-3mu_{2}$ system. Choosing the $z$-basis, the projection operator $\hat{X}_{+^{5}\mkern-5mu/\mkern-3mu_{2}}$ corresponding to the experimental atomic proposition ``The spin of the system along the $x$-axis is $+^{5}\mkern-5mu/\mkern-3mu_{2}\hbar\mkern3mu$'', denoted by $X_{+^{5}\mkern-5mu/\mkern-3mu_{2}}$, can be expressed in matrix form as follows\smallskip

\begin{equation}  
   \mathbf{M}(\hat{X}_{+^{5}\mkern-5mu/\mkern-3mu_{2}})
   =
   \frac{1}{32}
   \left[
      \begingroup
      \begin{array}{r r r r r r}
         1
         &
         \sqrt{5}
         &
         \sqrt{10}
         &
         \sqrt{10}
         &
         \sqrt{5}
         &
         1
         \\ 
         \sqrt{5}
         &
         5
         &
         5\sqrt{2}
         &
         5\sqrt{2}
         &
         5
         &
         \sqrt{5}
         \\ 
         \sqrt{10}
         &
         5\sqrt{2}
         &
         10
         &
         10
         &
         5\sqrt{2}
         &
         \sqrt{10}
         \\ 
         \sqrt{10}
         &
         5\sqrt{2}
         &
         10
         &
         10
         &
         5\sqrt{2}
         &
         \sqrt{10}
          \\ 
         \sqrt{5}
         &
         5
         &
         5\sqrt{2}
         &
         5\sqrt{2}
         &
         5
         &
         \sqrt{5}
         \\ 
         1
         &
         \sqrt{5}
         &
         \sqrt{10}
         &
         \sqrt{10}
         &
         \sqrt{5}
         &
         1 
     \end{array}
      \endgroup
   \right]
   \;\;\;\;  ;
\end{equation}
\smallskip

\noindent its range and kernel are\smallskip

\begin{equation}  
   \mathrm{ran}(\mathbf{M}(\hat{X}_{+^{5}\mkern-5mu/\mkern-3mu_{2}}))
   =
   \left\{
   a
   \left[
      \mkern-8mu
      \begingroup
      \begin{array}{r}
         1
         \\
         \sqrt{5}
         \\
         \sqrt{10}
         \\
         \sqrt{10}
          \\
         \sqrt{5}
         \\
         1
     \end{array}
      \endgroup
      \mkern-5mu
   \right]
   \right\}
   \;\;\;\;  ,
\end{equation}
\\[-35pt]

\begin{equation}  
   \mathrm{ker}(\mathbf{M}(\hat{X}_{+^{5}\mkern-5mu/\mkern-3mu_{2}}))
   =
   \left\{
   \mkern-3mu
   b_1
   \mkern-5mu
   \left[
      \mkern-8mu
      \begingroup
      \begin{array}{r}
         31
         \\
         -\sqrt{5}
         \\
         -\sqrt{10}
         \\
         -\sqrt{10}
          \\
         -\sqrt{5}
         \\
         -1
     \end{array}
      \mkern-5mu
      \endgroup
   \right]
   \mkern-5mu
   +
   b_2
   \mkern-5mu
   \left[
      \mkern-8mu
      \begingroup
      \begin{array}{r}
         -\sqrt{5}
         \\
         27
         \\
         -5\sqrt{2}
         \\
         -5\sqrt{2}
          \\
         -5
         \\
         -\sqrt{5}
     \end{array}
      \mkern-5mu
      \endgroup
   \right]
   \mkern-5mu
   +
   b_3
   \mkern-5mu
   \left[
      \mkern-8mu
      \begingroup
      \begin{array}{r}
         -\sqrt{10}
         \\
         -5\sqrt{2}
         \\
         22
         \\
         -10
          \\
         -5\sqrt{2}
         \\
         -\sqrt{10}
     \end{array}
      \endgroup
      \mkern-5mu
   \right]
   \mkern-5mu
   +
   b_4
   \mkern-5mu
   \left[
      \mkern-8mu
      \begingroup
      \begin{array}{r}
         -\sqrt{10}
         \\
         -5\sqrt{2}
         \\
         -10
         \\
         22
          \\
         -5\sqrt{2}
         \\
         -\sqrt{10}
     \end{array}
      \endgroup
      \mkern-5mu
   \right]
   \mkern-5mu
   +
   b_5
   \mkern-5mu
   \left[
       \mkern-8mu
     \begingroup
      \begin{array}{r}
         -\sqrt{5}
         \\
         -5
         \\
         -5\sqrt{2}
         \\
         -5\sqrt{2}
          \\
         27
         \\
         -\sqrt{5}
     \end{array}
      \endgroup
      \mkern-5mu
   \right]
   \mkern-5mu
   \right\}
   \;\;\;\;  ,
\end{equation}
\smallskip

\noindent where $a,b_1,…,b_5\in\mathbb{C}$. Suppose that this system is in the pure state described by the unit vector $|\Psi_{^{5}\mkern-5mu/\mkern-3mu_{2}}\rangle$ expressed as the column vector $\mathbf{\Psi}\in\mkern-2mu\mathbb{C}^{{6}\times{1}}$ with respect to the $z$-basis, namely,\smallskip

\begin{equation}  
   \mathbf{\Psi}
   =
   \mathbf{M}(|\Psi_{^{5}\mkern-5mu/\mkern-3mu_{2}}\rangle)
   =
   \frac{1}{4\sqrt{2}}
   \mkern-3mu
   \left[
      \mkern-5mu
      \begingroup
      \begin{array}{r r r r r r}
         \sqrt{5}
         &
         -3
         &
         \sqrt{2}
         &
         \sqrt{2}
         &
         -3
         &
         \sqrt{5}
     \end{array}
      \mkern-5mu
      \endgroup
   \right]^{\mathrm{T}}
   \;\;\;\;  .
\end{equation}
\smallskip

\noindent To decide which one of the statements – $|\Psi_{^{5}\mkern-5mu/\mkern-3mu_{2}}\rangle\mkern-2.5mu\in\mkern-2mu\mathrm{ran}(\hat{X}_{+^{5}\mkern-5mu/\mkern-3mu_{2}})$ or $|\Psi_{^{5}\mkern-5mu/\mkern-3mu_{2}}\rangle\mkern-2.5mu\in\mkern-2mu\mathrm{ker}(\hat{X}_{+^{5}\mkern-5mu/\mkern-3mu_{2}})$ – is true, let us present them as the systems of linear equations $\mathbf{RX}\mkern-3mu=\mkern-2mu\mathbf{\Psi}$ and $\mathbf{KX}\mkern-3mu=\mkern-2mu\mathbf{\Psi}$, correspondingly,\smallskip

\begin{equation}  
   \left[
      \mkern-8mu
      \begingroup
      \begin{array}{r}
         1
         \\
         \sqrt{5}
         \\
         \sqrt{10}
         \\
         \sqrt{10}
          \\
         \sqrt{5}
         \\
         1
     \end{array}
      \endgroup
      \mkern-5mu
   \right]
   x
   =
   \frac{1}{4\sqrt{2}}
   \mkern-3mu
   \left[
      \mkern-5mu
      \begingroup
      \begin{array}{r}
         \sqrt{5}
         \\
         -3
         \\
         \sqrt{2}
         \\
         \sqrt{2}
         \\
         -3
         \\
         \sqrt{5}
     \end{array}
      \mkern-5mu
      \endgroup
   \right]
   \;\;\;\;  ,
\end{equation}
\\[-25pt]

\begin{equation}  
   \left[
      \mkern-8mu
      \begingroup
      \begin{array}{r r r r r}
         31
         &
         -\sqrt{5}
         &
         -\sqrt{10}
         &
         -\sqrt{10}
         &
         -\sqrt{5}
         \\ 
         -\sqrt{5}
         &
         27
         &
         -5\sqrt{2}
         &
         -5\sqrt{2}
         &
         -5
         \\ 
         -\sqrt{10}
         &
         -5\sqrt{2}
         &
         22
         &
         -10
         &
         -5\sqrt{2}
         \\ 
         -\sqrt{10}
         &
         -5\sqrt{2}
         &
         -10
         &
         22
         &
         -5\sqrt{2}
         \\ 
         -\sqrt{5}
         &
         -5
         &
         -5\sqrt{2}
         &
         -5\sqrt{2}
         &
         27
         \\ 
         -1
         &
         -\sqrt{5}
         &
         -\sqrt{10}
         &
         -\sqrt{10}
         &
         -\sqrt{5}
     \end{array}
      \mkern-5mu
      \endgroup
   \right]
   \left[
      \mkern-5mu
      \begingroup
      \begin{array}{r}
         x_1
         \\
         x_2
         \\
         x_3
         \\
         x_4
         \\
         x_5
     \end{array}
      \mkern-5mu
      \endgroup
   \right]
   =
   \frac{1}{4\sqrt{2}}
   \mkern-3mu
   \left[
      \mkern-5mu
      \begingroup
      \begin{array}{r}
         \sqrt{5}
         \\
         -3
         \\
         \sqrt{2}
         \\
         \sqrt{2}
         \\
         -3
         \\
         \sqrt{5}
     \end{array}
      \mkern-5mu
      \endgroup
   \right]
   \;\;\;\;  .
\end{equation}
\smallskip

\noindent Even though both systems are overdetermined, the second one has the solution:\smallskip

\begin{equation}  
   \mathbf{X}
   =
   \frac{1}{32}
   \mkern-3mu
   \left[
      \mkern-5mu
      \begingroup
      \begin{array}{r r r r r r}
         0
         &
         -\sqrt{2}
         &
         -1
         &
         -1
         &
         -\sqrt{2}
     \end{array}
      \mkern-5mu
      \endgroup
   \right]^{\mathrm{T}}
   \;\;\;\;  .
\end{equation}
\smallskip

\noindent Hence, it is the case that the set $U_{\mathbf{K}}$ is not empty and thus $|\Psi_{^{5}\mkern-5mu/\mkern-3mu_{2}}\rangle$ belongs to $\mathrm{ker}(\hat{X}_{+^{5}\mkern-5mu/\mkern-3mu_{2}})$. According to (\ref{SV1}), this means that in the state $|\Psi_{^{5}\mkern-5mu/\mkern-3mu_{2}}\rangle$ the proposition $X_{+^{5}\mkern-5mu/\mkern-3mu_{2}}$ is false.\\

\section{Conjecture of equal cost of computation}  

\noindent First thing to notice in the formula (\ref{SV1}), the assignment of truth values to an experimental atomic proposition $P$ in a given pure quantum state $|\Psi\rangle$ can be treated as a computational problem, namely, the inspection of solvability of the systems of linear equations $\mathbf{RX}\mkern-3mu=\mkern-2mu\mathbf{\Psi}$ and $\mathbf{KX}\mkern-3mu=\mkern-2mu\mathbf{\Psi}$.\\

\noindent Let $[\cdot]_{\mathbf{R}}$ denote a metric used to gauge the performance of the algorithm for verifying the statement $U_{\mathbf{R}}\neq\varnothing$. Similarly, let $[\cdot]_{\mathbf{K}}$ stand for a metric evaluating the performance of the algorithm for verifying the statement $U_{\mathbf{K}}\neq\varnothing$, and let $[\cdot]_{\varnothing}$ refer to a metric used to estimate the performance of the algorithm for verifying $U_{\mathbf{R}}\mkern-2mu=\mkern-2mu\varnothing{\mkern3mu\sqcap}\mkern3muU_{\mathbf{K}}\mkern-2mu=\mkern-2mu\varnothing$.\\

\noindent Let us define the work of a computation, $W$, as the total number of primitive operations performed to solve the computational problem at hand. Also, let us define the cost of the computation, $C$, as the amount of time taken to solve this problem. Providing each primitive operation takes some fixed amount of time to perform, the work $W$ and the cost $C$ can be compared with each other (for example, they can be equal or unequal depending on a model of computation).\\

\noindent Regarding the formula (\ref{SV1}), one can assume the following:
\\[-10pt]

\begin{conjecture}  
\noindent For any experimental atomic proposition referring to a quantum system in a pure state, the cost of the computation of a truth value does not depend on the truth value itself or on its absence.
\end{conjecture}

\noindent In accordance with the proposed notation, $[C]_{\mathbf{R}}$ denotes the cost of the computation invested in verifying that the solution set $U_{\mathbf{R}}$ is not empty, $[C]_{\mathbf{K}}$ stands for the cost of the computation executed to check that the set $U_{\mathbf{K}}$ has at least one solution, and $[C]_{\varnothing}$ is the cost of the computation needed to confirm that those sets are empty. Conjecture 1, if true, implies:\smallskip

\begin{equation} \label{EQV} 
   [C]_{\mathbf{R}}
   =
   [C]_{\mathbf{K}}
   =
   [C]_{\varnothing}
   \;\;\;\;  ,
\end{equation}
\smallskip

\noindent where the sign ``$=$'' means to express that $[C]_{\mathbf{R}}$, $[C]_{\mathbf{K}}$ and $ [C]_{\varnothing}$ – as functions of the dimension $n$ of the Hilbert space $\mathcal{H}$ – have equal growth rates.\\

\noindent The problem is that the equivalence (\ref{EQV}) cannot be achieved on a RAM (Random Access Machine), i.e., a model of sequential computation. To demonstrate this, let us define numbers of primitive operations required to verify that the sets $U_{\mathbf{R}}$ and $U_{\mathbf{K}}$ are not empty.\\

\noindent The linear system $\mathbf{RX}\mkern-3mu=\mkern-2mu\mathbf{\Psi}$ has only one unknown, so, being presented as an augmented matrix it takes the form:\smallskip

\begin{equation}  
   \left[
   \mathbf{R}
   \mkern-2mu
   \left|
   \mathbf{\Psi}
   \right.
   \right]
   \equiv
   \left[
      \mkern-8mu
      \begingroup
      \begin{array}{c | c}
         P_{11}
         &
         b_1
        \\
         P_{21}
         &
         b_2
         \\
         P_{31}
         &
         b_3
         \\
         \vdots
         &
         \vdots
         \\
         P_{n.1}
         &
         b_n
      \end{array}
      \endgroup
      \mkern-8mu
    \right]
  \;\;\;\;  ,
\end{equation}
\smallskip

\noindent where $b_j=\langle{e_j}|\Psi\rangle$. This system will be consistent if the following condition holds:\smallskip

\begin{equation} \label{COMP} 
   \forall
   j
   \in
   \left\{
      2,
      \dots
      ,
      n
   \right\}
   :
   \mkern8mu
   b_1
   P_{j.1}
   =
   b_j
   P_{11}
   \;\;\;\;  .
\end{equation}
\smallskip

\noindent Subsequently, to confirm that the set $U_{\mathbf{R}}$ is not empty requires $2(n-1)$ multiplications and $n-1$ comparisons (note that comparisons as well as elementary arithmetic operations can be regarded as primitive operations). Even though a comparison between real or complex numbers with infinite precision may not be realistic, such a procedure (along with any elementary arithmetic operation on real or complex numbers) can be seen as an approximation to the asymptotic behavior as more and more precision is allowed to use in the computation.\\

\noindent Obviously, to prove that the set $U_{\mathbf{R}}$ is empty will require the same number of multiplications and, in the worst case, $n-2$ comparisons.\\

\noindent The linear system of equations $\mathbf{KX}\mkern-3mu=\mkern-2mu\mathbf{\Psi}$ has $n-1$ unknowns, so its augmented matrix is\smallskip

\begin{equation}  
   \left[
   \mathbf{K}
   \mkern-2mu
   \left|
   \mathbf{\Psi}
   \right.
   \right]
   \equiv
   \left[
      \mkern-8mu
      \begingroup
      \begin{array}{c c c c c | c}
         a_{11}
         &
         a_{12}
         &
         a_{13}
         &
         \cdots
         &
         a_{1.n-1}
         &
         a_{1.n}
         \\ 
         a_{21}
         &
         a_{22}
         &
         a_{23}
         &
         \cdots
         &
         a_{2.n-1}
         &
         a_{2.n}
         \\ 
         a_{31}
         &
         a_{32}
         &
         a_{33}
         &
         \cdots
         &
         a_{3.n-1}
         &
         a_{3.n}
         \\ 
         \vdots
         &
         \vdots
         &
         \vdots
         &
         \ddots
         &
         \vdots
         &
         \vdots
         \\ 
         a_{n.1}
         &
         a_{n.2}
         &
         a_{n.3}
         &
         \cdots
         &
         a_{n.n-1}
         &
         a_{n.n}
     \end{array}
      \mkern-5mu
      \endgroup
   \right]
   \;\;\;\;  ,
\end{equation}
\smallskip

\noindent where\smallskip

\begin{equation} 
   a_{jl}
   =
   \left\{
      \begingroup\SmallColSep
      \begin{array}{r l}
         \delta_{jl}
         -
         P_{jl}
         ,
         &
         \mkern15mu
         l
         <
         n
         \\[5pt]
         \langle{e_j}|\Psi\rangle
         ,
         &
         \mkern15mu
         l
         =
         n
      \end{array}
      \endgroup
   \right.
   \;\;\;\;  .
\end{equation}
\smallskip

\noindent To bring this matrix to the reduced row echelon form

\begin{equation}  
   \left[
      \mkern-8mu
      \begingroup
      \begin{array}{c c c c c | c}
         a_{11}
         &
         a_{12}
         &
         a_{13}
         &
         \cdots
         &
         a_{1.n-1}
         &
         a_{1.n}
         \\[3pt] 
         0
         &
         a_{22}^{(1)}
         &
         a_{23}^{(1)}
         &
         \cdots
         &
         a_{2.n-1}^{(1)}
         &
         a_{2.n}^{(1)}
         \\[3pt] 
         0
         &
         0
         &
         a_{33}^{(2)}
         &
         \cdots
         &
         a_{3.n-1}^{(2)}
         &
         a_{3.n}^{(2)}
         \\ 
         \vdots
         &
         \vdots
         &
         \vdots
         &
         \ddots
         &
         \vdots
         &
         \vdots
         \\[3pt] 
         0
         &
         0
         &
         0
         &
         \cdots
         &
         a_{n-1.n-1}^{(n-2)}
         &
         a_{n-1.n}^{(n-2)}
         \\[3pt] 
         0
         &
         0
         &
         0
         &
         \cdots
         &
         a_{n.n-1}^{(n-2)}
         &
         a_{n.n}^{(n-2)}
     \end{array}
      \mkern-5mu
      \endgroup
   \right]
   \;\;\;\;  ,
\end{equation}
\smallskip

\noindent requires application of the iteration formula\smallskip

\begin{equation} \label{GAUSS}
   \begingroup\SmallColSep
   \begin{array}{r}
      i
      \in
      \left\{
         1
         ,
         \dots
         ,
         n-2
      \right\}
      \\
      j
      \in
      \left\{
         i+1
         ,
         i+2
         ,
         \dots
         n
      \right\}
      \\
      l
      \in
      \left\{
         i+1
         ,
         i+2
         ,
         \dots
         n
      \right\}
   \end{array}
   \endgroup
   \mkern-2mu
   :
   \mkern10mu
   a_{jl}^{(i)}
   =
   a_{jl}^{(i-1)}
   -
   \frac{a_{ji}^{(i-1)}}{a_{ii}^{(i-1)}}
   a_{il}^{(i-1)}
   \;\;\;\;  ,
\end{equation}
\smallskip

\noindent which executes\smallskip

\begin{equation} 
   \sum_{i=1}^{n-2}
      (n-1)
   =
   \frac{n(n-1)}{2}
   -
   1
   \;\;\;\;  
\end{equation}
\smallskip

\noindent divisions along with\smallskip

\begin{equation} 
   \sum_{i=1}^{n-2}
      (n-1)^2
   =
   \frac{n(n-1)(2n-1)}{6}
   -
   1
   \;\;\;\;  
\end{equation}
\smallskip

\noindent multiplications and the same number of subtractions. Once in the row echelon form, the condition\smallskip

\begin{equation} \label{LCOND} 
   a_{n.n-1}^{(n-2)}
   \cdot
   a_{n-1.n}^{(n-2)}
   =
   a_{n-1.n-1}^{(n-2)}
   \cdot
   a_{n.n}^{(n-2)}
   \;\;\;\;  
\end{equation}
\smallskip

\noindent should be examined and if it holds, the system $\mathbf{KX}\mkern-3mu=\mkern-2mu\mathbf{\Psi}$ is consistent and at least one solution exists; otherwise, the solution set $U_{\mathbf{K}}$ is empty.\\

\noindent Note that the formula (\ref{GAUSS}) breaks down if $a_{ii}^{(i-1)}=0$. This can be avoided by choosing the largest (in absolute value) of $a_{ii}^{(i-1)},a_{i+1.i}^{(i-1)},\dots,a_{n-2.i}^{(i-1)}$ and interchanging its row with the $i^{\text{th}}$  row before applying the formula (\ref{GAUSS}). Such a procedure is known as Gauss's method with partial pivoting. But even so, the work of this procedure will remain the same, that is, $O(n^3)$ \cite{Robert}.\\

\noindent In the RAM model, primitive operations are executed one after another, so the work can be considered equal to the cost (to be exact, $W$ and $C$ differ by at most a constant factor) \cite{Cormen}. One can infer from here that $[C]_{\mathbf{R}}=O(n)$ whereas $[C]_{\mathbf{K}}=[C]_{\varnothing}=O(n^3)$.\\

\noindent Thus, serial computation contradicts Conjecture 1.\\
 
\noindent This conclusion assumes that the presented above algorithm for the confirmation of the statement $U_{\mathbf{K}}\neq\varnothing$ uses asymptotically optimal number of primitive operations. It means that, for larger enough dimensions $n$ of Hilbert spaces, no algorithm verifying the said statement can have a number of primitive operations growing significantly slower than $O(n^3)$.\\

\noindent Let us consider a matrix version of the algorithm checking the solvability of the system $\mathbf{KX}\mkern-3mu=\mkern-2mu\mathbf{\Psi}$.\\

\noindent At the first step of the algorithm, on condition that $a_{11}\neq0$, divide the first column vector of the matrix $\left[\mathbf{K}\mkern-2mu\left|\mathbf{\Psi}\right.\right]$ by $a_{11}$, explicitly,\smallskip

\begin{equation}  
   \left[
      \mkern-8mu
      \begingroup
      \begin{array}{c}
         a_{11}
        \\
         a_{21}
         \\
         a_{31}
         \\
         \vdots
         \\
         a_{n.1}
      \end{array}
      \endgroup
      \mkern-8mu
   \right]
   \div
   a_{11}
   =
   \left[
      \mkern-8mu
      \begingroup
      \begin{array}{c}
         1
        \\
         c_{21}
         \\
         c_{31}
         \\
         \vdots
         \\
         c_{n.1}
      \end{array}
      \endgroup
      \mkern-8mu
   \right]
   \;\;\;\;  ,
\end{equation}
\smallskip

\noindent where $c_{j1}$ denotes $a_{j1}/a_{11}$. Find the outer product of the resultant column vector and the first row vector of $\left[\mathbf{K}\mkern-2mu\left|\mathbf{\Psi}\right.\right]$:\smallskip

\begin{equation}  
   \mathbf{D}
   =
   \left[
      \mkern-8mu
      \begingroup
      \begin{array}{c}
         1
         \\
         c_{21}
         \\
         c_{31}
         \\
         \vdots
         \\
         c_{n.1}
      \end{array}
      \endgroup
      \mkern-8mu
   \right]
   \mkern-2mu
   \left[
      \begingroup\SmallColSep
      \begin{array}{c c c c c c}
         a_{11}
         \mkern3mu
         &
         a_{12}
         \mkern3mu
         &
         a_{13}
         \mkern3mu
         &
         \cdots
         &
         a_{1.n-1}
         \mkern3mu
         &
         a_{1.n}
         \mkern-1.5mu
      \end{array}
      \endgroup
   \right]
   \mkern-2mu
   =
   \mkern-2mu
   \left[
      \mkern-8mu
      \begingroup
      \begin{array}{c c c c c | c}
         a_{11}
         &
         a_{12}
         &
         a_{13}
         &
         \cdots
         &
         a_{1.n-1}
         &
         a_{1.n}
         \\ 
         a_{21}
         &
         d_{22}
         &
         d_{23}
         &
         \cdots
         &
         d_{2.n-1}
         &
         d_{2.n}
         \\ 
         a_{31}
         &
         d_{32}
         &
         d_{33}
         &
         \cdots
         &
         d_{3.n-1}
         &
         d_{3.n}
         \\ 
         \vdots
         &
         \vdots
         &
         \vdots
         &
         \ddots
         &
         \vdots
         &
         \vdots
         \\ 
         a_{n.1}
         &
         d_{n.2}
         &
         d_{n.3}
         &
         \cdots
         &
         d_{n.n-1}
         &
         d_{n.n}
      \end{array}
      \endgroup
      \mkern-5mu
   \right]
   \;\;\;\;  ,
\end{equation}
\smallskip

\noindent where $d_{jl}=c_{j1}a_{1l}$. Now, take this product from $\left[\mathbf{K}\mkern-2mu\left|\mathbf{\Psi}\right.\right]$:\\

\begin{equation}  
   \left[\mathbf{K}\mkern-2mu\left|\mathbf{\Psi}\right.\right]^{(1)}
   =
   \left[\mathbf{K}\mkern-2mu\left|\mathbf{\Psi}\right.\right]
   -
   \mathbf{D}
   =
   \mkern-2mu
   \left[
      \mkern-8mu
      \begingroup
      \begin{array}{c c c c c | c}
         0
         &
         0
         &
         0
         &
         \cdots
         &
         0
         &
         0
         \\[3pt]  
         0
         &
         a_{22}^{(1)}
         &
         a_{23}^{(1)}
         &
         \cdots
         &
         a_{2.n-1}^{(1)}
         &
         a_{2.n}^{(1)}
         \\[3pt]  
         0
         &
         a_{32}^{(1)}
         &
         a_{33}^{(1)}
         &
         \cdots
         &
         a_{3.n-1}^{(1)}
         &
         a_{3.n}^{(1)}
         \\ 
         \vdots
         &
         \vdots
         &
         \vdots
         &
         \ddots
         &
         \vdots
         &
         \vdots
         \\[3pt]  
         0
         &
         a_{n.2}^{(1)}
         &
         a_{n.3}^{(1)}
         &
         \cdots
         &
         a_{n.n-1}^{(1)}
         &
         a_{n.n}^{(1)}
      \end{array}
      \endgroup
      \mkern-5mu
   \right]
   \;\;\;\;  ,
\end{equation}
\smallskip

\noindent where $a_{jl}^{(1)}=a_{jl}-c_{j1}a_{1l}$.\\

\noindent Similarly, at the second step of the algorithm, if $a_{22}^{(1)}\neq0$, divide the second column vector of the matrix $\left[\mathbf{K}\mkern-2mu\left|\mathbf{\Psi}\right.\right]^{(1)}$ by $a_{22}^{(1)}$, i.e.,\smallskip

\begin{equation}  
   \left[
      \mkern-8mu
      \begingroup
      \begin{array}{c}
         0
         \\[3pt]
         a_{22}^{(1)}
         \\[3pt]
         a_{32}^{(1)}
         \\
         \vdots
         \\[3pt]
         a_{n.2}^{(1)}
      \end{array}
      \endgroup
      \mkern-8mu
   \right]
   \div
   a_{22}^{(1)}
   =
   \left[
      \mkern-8mu
      \begingroup
      \begin{array}{c}
         0
        \\
         1
         \\
         c_{32}
         \\
         \vdots
         \\
         c_{n.2}
      \end{array}
      \endgroup
      \mkern-8mu
   \right]
   \;\;\;\;  ,
\end{equation}
\smallskip

\noindent where $c_{j2}=a_{j2}^{(1)}/a_{22}^{(1)}$. Find the outer product of the resultant vector and the second row vector of $\left[\mathbf{K}\mkern-2mu\left|\mathbf{\Psi}\right.\right]^{(1)}$, that is,\smallskip

\begin{equation}  
   \mathbf{D}^{(1)}
   =
   \left[
      \mkern-8mu
      \begingroup
      \begin{array}{c}
         0
         \\
         1
         \\
         c_{32}
         \\
         \vdots
         \\
         c_{n.2}
      \end{array}
      \endgroup
      \mkern-8mu
   \right]
   \mkern-2mu
   \left[
      \begingroup\SmallColSep
      \begin{array}{c c c c c c}
         0
         &
         a_{22}^{(1)}
         \mkern3mu
         &
         a_{23}^{(1)}
         \mkern3mu
         &
         \cdots
         &
         a_{2.n-1}^{(1)}
         \mkern3mu
         &
         a_{2.n}^{(1)}
         \mkern-1.5mu
      \end{array}
      \endgroup
   \right]
   \mkern-2mu
   =
   \mkern-2mu
   \left[
      \mkern-8mu
      \begingroup
      \begin{array}{c c c c c | c}
         0
         &
         0
         &
         0
         &
         \cdots
         &
         0
         &
         0
         \\[3pt]  
         0
         &
         a_{22}^{(1)}
         &
         a_{23}^{(1)}
         &
         \cdots
         &
         a_{2.n-1}^{(1)}
         &
         a_{2.n}^{(1)}
         \\[3pt]  
         0
         &
         a_{32}^{(1)}
         &
         d_{33}^{\mkern1.5mu(1)}
         &
         \cdots
         &
         d_{3.n-1}^{\mkern1.5mu(1)}
         &
         d_{3.n}^{\mkern1.5mu(1)}
         \\ 
         \vdots
         &
         \vdots
         &
         \vdots
         &
         \ddots
         &
         \vdots
         &
         \vdots
         \\[3pt]  
         0
         &
         a_{n.2}^{(1)}
         &
         d_{n.3}^{\mkern1.5mu(1)}
         &
         \cdots
         &
         d_{n.n-1}^{\mkern1.5mu(1)}
         &
         d_{n.n}^{\mkern1.5mu(1)}
      \end{array}
      \endgroup
      \mkern-5mu
   \right]
   \;\;\;\;  ,
\end{equation}
\smallskip

\noindent where $d_{jl}^{\mkern1.5mu(1)}=c_{j2}a_{2l}^{(1)}$, and afterwards subtract this product from $\left[\mathbf{K}\mkern-2mu\left|\mathbf{\Psi}\right.\right]^{(1)}$:\smallskip

\begin{equation}  
   \left[\mathbf{K}\mkern-2mu\left|\mathbf{\Psi}\right.\right]^{(2)}
   =
   \left[\mathbf{K}\mkern-2mu\left|\mathbf{\Psi}\right.\right]^{(1)}
   -
   \mathbf{D}^{(1)}
   =
   \mkern-2mu
   \left[
      \mkern-8mu
      \begingroup
      \begin{array}{c c c c c | c}
         0
         &
         0
         &
         0
         &
         \cdots
         &
         0
         &
         0
         \\[3pt]  
         0
         &
         0
         &
         0
         &
         \cdots
         &
         0
         &
         0
         \\[3pt]  
         0
         &
         0
         &
         a_{33}^{(2)}
         &
         \cdots
         &
         a_{3.n-1}^{(2)}
         &
         a_{3.n}^{(2)}
         \\ 
         \vdots
         &
         \vdots
         &
         \vdots
         &
         \ddots
         &
         \vdots
         &
         \vdots
         \\[3pt]  
         0
         &
         0
         &
         a_{n.3}^{(2)}
         &
         \cdots
         &
         a_{n.n-1}^{(2)}
         &
         a_{n.n}^{(2)}
      \end{array}
      \endgroup
      \mkern-5mu
   \right]
   \;\;\;\;  ,
\end{equation}
\smallskip

\noindent where $a_{jl}^{(2)}=a_{jl}^{(1)}-c_{j2}a_{2l}^{(1)}$.\\

\noindent Subsequently, at the $i^{\text{th}}$ step of the algorithm, the matrix $\left[\mathbf{K}\mkern-2mu\left|\mathbf{\Psi}\right.\right]^{(i)}$ will be computed by the formula\smallskip

\begin{equation}  
   \left[\mathbf{K}\mkern-2mu\left|\mathbf{\Psi}\right.\right]^{(i)}
   =
   \left[\mathbf{K}\mkern-2mu\left|\mathbf{\Psi}\right.\right]^{(i-1)}
   -
   \mathbf{D}^{(i-1)}
   \;\;\;\;  ,
\end{equation}
\smallskip

\noindent where\smallskip

\begin{equation}  
   \mathbf{D}^{(i-1)}
   =
   \left(
      \mkern-3mu
      \left[
         \mkern-8mu
         \begingroup
         \begin{array}{r}
            (i-1)
            \left\{
               \mkern-2mu
               \begingroup\SmallColSep
               \begin{array}{r}
                  0
                  \\
                  \vdots
                  \\
                  0
               \end{array}
               \endgroup
            \right.
            \\[19pt]
            a_{ii}^{(i-1)}
            \\[3pt]
            a_{i+1.i}^{(i-1)}
            \\
            \vdots
            \\[3pt]
            a_{n.i}^{(i-1)}
         \end{array}
         \endgroup
         \mkern-8mu
      \right]
      \mkern-2mu
      \div
      \mkern-2mu
      a_{ii}^{(i-1)}
      \mkern-3mu
   \right)
   \left[
      \underbrace{0\mkern3mu\cdots\mkern3mu0}_{\text{\normalsize{$(i-1)$}}}
      \begingroup\SmallColSep
      \begin{array}{c c c c}
         \mkern3mu
         a_{ii}^{(i-1)}
         \mkern3mu
         &
         a_{i.i+1}^{(i-1)}
         \mkern3mu
         &
         \cdots
         &
         a_{i.n}^{(i-1)}
      \end{array}
      \endgroup
      \mkern-5mu
   \right]
   \;\;\;\;   
\end{equation}
\smallskip

\noindent as long as $a_{ii}^{(i-1)}\neq0$.\\

\noindent Eventually, one gets the matrix\smallskip

\begin{equation}  
   \left[\mathbf{K}\mkern-2mu\left|\mathbf{\Psi}\right.\right]^{(n-2)}
   =
   \mkern-2mu
   \left[
      \mkern-8mu
      \begingroup
      \begin{array}{c c c c c | c}
         0
         &
         0
         &
         0
         &
         \cdots
         &
         0
         &
         0
         \\[3pt]  
         0
         &
         0
         &
         0
         &
         \cdots
         &
         0
         &
         0
         \\[3pt]  
         0
         &
         0
         &
         0
         &
         \cdots
         &
         0
         &
         0
         \\ 
         \vdots
         &
         \vdots
         &
         \vdots
         &
         \ddots
         &
         \vdots
         &
         \vdots
         \\[3pt]  
         0
         &
         0
         &
         0
         &
         \cdots
         &
         a_{n-1.n-1}^{(n-2)}
         &
         a_{n-1.n}^{(n-2)}
         \\[3pt]  
         0
         &
         0
         &
         0
         &
         \cdots
         &
         a_{n.n-1}^{(n-2)}
         &
         a_{n.n}^{(n-2)}
      \end{array}
      \endgroup
      \mkern-5mu
   \right]
   \;\;\;\;  ,
\end{equation}
\smallskip

\noindent whose nonzero entries are $a_{jl}^{(n-2)}=a_{jl}^{(n-1)}-c_{j.n-2}\cdot a_{n-2.l}^{(n-1)}$. If these entries meet the condition (\ref{LCOND}), then the set $U_{\mathbf{K}}$ is not empty.\\

\noindent To compute $\left[\mathbf{K}\mkern-2mu\left|\mathbf{\Psi}\right.\right]^{(i)}$ takes $O(n)$ divisions, $O(n^2)$ multiplications and the same number of subtractions. Subsequently, the cost of the computation of $\left[\mathbf{K}\mkern-2mu\left|\mathbf{\Psi}\right.\right]^{(i)}$ on a RAM will be $O(n^2)$.\\

\noindent However, to enable Conjecture 1, the cost $[C]_{\mathbf{K}}$ must grow no faster than $[C]_{\mathbf{R}}$. Given that the algorithm for the confirmation of the statement $U_{\mathbf{K}}\neq\varnothing$ has $n-2=O(n)$ steps (performed sequentially because of data dependencies), this implies that the cost of the computation of $\left[\mathbf{K}\mkern-2mu\left|\mathbf{\Psi}\right.\right]^{(i)}$ must be constant, i.e., $O(1)$.\\

\noindent Inasmuch as this is impossible with serial computing, one may hope to achieve this using parallel computing, i.e., by executing operations in parallel.\\

\section{Conjecture of quantum parallel computing}  

\noindent Note that the algorithm for the confirmation of the statement $U_{\mathbf{R}}\neq\varnothing$ relies on $O(n)$ comparisons performed in sequence: see (\ref{COMP}). Particularly, each such comparison (except the first one) requires the positive answer from the preceding comparison in order to be executed. Thus, one may reasonably expect that the said algorithm cannot significantly be sped up by using parallelism.\\

\noindent In contrast, the matrix version of the algorithm for the confirmation of the statement $U_{\mathbf{K}}\neq\varnothing$ is split up into $O(n)$ steps in a manner that each step, though being performed in sequence, contains only operations which are easily parallelizable, namely, scalar division, outer product and matrix subtraction \cite{Cormen}.\\

\noindent Consider a classical parallel RAM (PRAM for short) that solves the computational problem at hand in time $T_p$ with $p$ classical processors. The cost of the computation of the problem on the PRAM is defined as $C_p = p \cdot T_p$. This quantity expresses the total time spent by $p$ processors in both computing and waiting \cite{Casanova, Reif}. The cost $C_p$ provides the lower bound on the running time $T_p$ of parallel computation. Explicitly, according to the work law,\smallskip

\begin{equation}  
   T_p
   \ge
   \frac{T_1}{p}
   \;\;\;\;  ,
\end{equation}
\smallskip

\noindent where $T_1$ stands for the time used to run the computation on a single processor. Ignoring the cost of the communication (attributable to the synchronization of the processors), the time $T_1$ is equal to the work $W$. Hence, with classical parallel computation, the cost is always at least the work, in symbols, $C_p \ge W$. This follows from the fact that the efficiency of classical parallel computing, denoted $E_p$, is not more than 1, explicitly,\smallskip

\begin{equation}  
   E_p
   =
   \frac{S_p}{p}
   \le
   1
   \;\;\;\;  ,
\end{equation}
\smallskip

\noindent where $S_p$ is the speedup\smallskip

\begin{equation}  
   S_p
   =
   \frac{T_1}{T_p}
   \;\;\;\;  ,
\end{equation}
\smallskip

\noindent i.e., the gain in speed made by parallel computing compared to sequential execution. Otherwise stated, the speedup on $p$ classical processors can be at most $p$.\\

\noindent The span of parallel computation denoted by $T_{\infty}$ is defined as the length of the longest series of operations that must be performed sequentially. On the other hand, the span $T_{\infty}$ can be viewed as the time spent computing on a PRAM with an unlimited number of processors. The ratio $T_1/T_{\infty}$, which is called the parallelism of the multithreaded computation, represents the maximum possible speedup on any number of processors. Indeed, as ${T_p}\ge{T_{\infty}}$, it holds\smallskip

\begin{equation}  
   \frac{T_1}{T_{\infty}}
   \ge
   \frac{T_1}{T_{p}}
   \;\;\;\;  .
\end{equation}
\smallskip

\noindent Since the computation of the matrix $\left[\mathbf{K}\mkern-2mu\left|\mathbf{\Psi}\right.\right]^{(i)}$ is easily parallelizable (specifically, the number of operations performed in sequence computing this matrix does not depend on $n$), its span can be regarded as constant, $T_{\infty}=O(1)$. This implies that the parallel computation of $\left[\mathbf{K}\mkern-2mu\left|\mathbf{\Psi}\right.\right]^{(i)}$ can be done in $O(1)$ time. But this does not mean that the cost of this computation can be $O(1)$ too.\\

\noindent To be sure, the number of processors required to achieve the perfect linear speedup (when the efficiency of the parallel computation comes to be maximal, i.e., 1) must be equal to the parallelism, so\smallskip

\begin{equation}  
   p
   =
   \frac{T_1}{T_{\infty}}
   =
   O(n^2)
   \;\;\;\;  .
\end{equation}
\smallskip

\noindent This makes the cost of the multithreaded computation of $\left[\mathbf{K}\mkern-2mu\left|\mathbf{\Psi}\right.\right]^{(i)}$ equal to\smallskip

\begin{equation}  
   C_p
   =
   O(n^2)
   O(1)
   =
   O(n^2)
   \;\;\;\;  .
\end{equation}
\smallskip

\noindent Concluding so far, neither serial computing nor classical parallel computing supports Conjecture 1.\\

\noindent Therefore, let us consider a special kind of a PRAM, namely, an abstract computational device, which uses quantum mechanics to perform parallel calculations \cite{Preskill, Shor}. Such a device is denoted quantum PRAM (QPRAM for short). According to the Quantum Parallelism Thesis, a QPRAM enjoys an advantage over a classical PRAM because it can invoke ``quantum parallelism''. This refers to a capacity to carry out many computations simultaneously in superposition of quantum states (see, for example, \cite{Timpson, Duwell2007, Horsman, Wallace, Lini, Duwell2018}). Due to the lack of a full understanding of what quantum parallelism exactly means (and why a QPRAM is superior to a classical PRAM \cite{Calude}), one can assume this:\\[-10pt]

\begin{conjecture}  
\noindent Unlike classical parallel computing, the efficiency of quantum parallel computing can be greater than 1.
\end{conjecture}

\noindent Let us denote the running time of a computation on $q$ quantum processors by $X_q$. Then, Conjecture 2 if true would imply that the speedup on $q$ quantum processors, defined as $S_q=T_1/X_q$, could be greater than $q$ (in this way, $E_q$, the efficiency of quantum parallel computing – i.e., the speedup per quantum processor – might be greater than 1; in symbols, $E_q=S_q/q\ge1$). This would yield the inequality\smallskip

\begin{equation} \label{QCOST} 
   q
   \cdot
   X_q
   \le
   T_1
   \;\;\;\;   
\end{equation}
\smallskip

\noindent meaning that $C_q = q \cdot X_q$, the cost of a computation on a QPRAM, can be less than the work $T_1$, the time taken to execute the entire computation sequentially, i.e., on a classical RAM. This  inequality might be viewed as a result of multipartite entanglement which arises from a QPRAM being made up of $q$ quantum processors \cite{Jozsa98, Jozsa02}.\\

\noindent It is reasonable to expect that a QPRAM with a finite number $q$ of processors cannot run any faster than a PRAM with an unlimited number of processors. Accordingly, the span law follows:\smallskip

\begin{equation} \label{SLAW} 
   X_q
   \ge
   T_{\infty}
   \;\;\;\;  .
\end{equation}
\smallskip

\noindent The span law implies that the classical parallelism bounds the quantum speedup:\smallskip

\begin{equation} \label{QSL} 
   \frac{T_1}{T_{\infty}}
   \ge
   \frac{T_1}{X_q}
   \ge
   q
   \;\;\;\;  .
\end{equation}
\smallskip

\noindent Suppose that $p=T_1/T_{\infty}$ and let $X_q$ be equal to $T_p$. Then, from the above formula it follows:\smallskip

\begin{equation}  
   p
   =
   \frac{T_1}{T_p}
   \ge
   q
   \;\;\;\;  ,
\end{equation}
\smallskip

\noindent and so\smallskip

\begin{equation}  
   C_p
   \ge
   C_q
   \;\;\;\;  .
\end{equation}
\smallskip

\noindent This signifies the quantum supremacy, namely, the cost of a computation on a QPRAM can be less than that on a PRAM.\\

\noindent For the computation of the matrix $\left[\mathbf{K}\mkern-2mu\left|\mathbf{\Psi}\right.\right]^{(i)}$ this means that it can be done in $O(1)$ time using $O(1)$ quantum processors. As a result, the total time spent computing all $O(n)$ steps of the algorithm verifying the statement $U_{\mathbf{K}}\neq\varnothing$ on a QPRAM can be $[C_q ]_{\mathbf{K}}=O(n)$.\\

\noindent On the other hand, since one cannot expect any gain in speed made by a parallel execution of the algorithm that verifies the statement $U_{\mathbf{R}}\neq\varnothing$, the parallelism of its multithreaded computation can be considered equal to $T_{1}/T_{\infty}=O(1)$. Substituting this in the formula (\ref{QSL}) produces\smallskip

\begin{equation}  
   O(1)
   =
   \left[
      \frac{T_1}{X_q}
   \right]_{\mathbf{R}}
   =
   \left[
      q
   \right]_{\mathbf{R}}
   \;\;\;\;  ,
\end{equation}
\smallskip

\noindent which involves $[C_q ]_{\mathbf{R}}=O(n)$ and, hence, the case that  $[C_q ]_{\mathbf{R}}=[C_q ]_{\mathbf{K}}$.\\

\noindent Consequently if Conjecture 2 held, then the equal cost of computations of the values of true and false for experimental quantum propositions could be possible.\\

\section{Discussion}  

\noindent Even though Conjecture 2 is open (it has not known yet how far capabilities of quantum computers are beyond those of conventional PRAMs \cite{Hagar, Resch}), let us suppose for a moment that Conjecture 2 had been disproved. Possible consequences of that might be as follows.\\

\noindent Since classical computing – sequential and parallel alike – cannot ensure the same cost of computation for both truth values of experimental quantum propositions, Conjecture 1 could not hold if Conjecture 2 were to be proven incorrect (and therewithal no additional supposition had been added).\\

\noindent The loss of Conjecture 1 would in turn entail a failure of a standard propositional calculus used to infer a conclusion from a premise to create an argument \cite{Brown, Klement}.\\

\noindent To be sure, consider, for example, the disjunction of propositions $P$ and $R$, i.e., $P \sqcup R$. Classically, if $P$ is true, then $P \sqcup R$ is true, no matter whether or not $R$ is true. Thus, the premise, i.e., the truth of $P$, implies the conclusion, i.e., the truth of $P \sqcup R$; in symbols:\smallskip

\begin{equation}  
   P
   \to
   P
   \sqcup
   R
   \;\;\;\;  .
\end{equation}
\smallskip

\noindent But, if Conjecture 1 were to be false, one would be able to differentiate the case, where both operands of $P \sqcup R$ were true, from the case, where only one of them was true, due to the fact that the costs of computations of such cases would be different. Namely, two conclusions, ``$P$ is true AND $R$ is true'' and ``$P$ is true AND $R$ is false'', would have the same value but the different costs of computations. For that reason, a one-to-one correspondence between the premise ``$P$ is true'' and those conclusions would be impossible. That is, the premise would no longer imply the conclusion.\\

\noindent To make Conjecture 1 possible without implying Conjecture 2 one can assume the following:\\[-10pt]

\begin{conjecture}  
\noindent Under the function $v$, the truth of the statement $|\Psi\rangle\notin\mathrm{ran}(\hat{P})\sqcap|\Psi\rangle\notin\mathrm{ker}(\hat{P})$ has the same image as the truth of $|\Psi\rangle\in\mathrm{ran}(\hat{P})$ or the truth of $|\Psi\rangle\in\mathrm{ker}(\hat{P})$ does.
\end{conjecture}

\noindent Conjecture 3, if true, would imply that\smallskip

\begin{equation}  
   v
   \left(
      |\Psi\rangle
      \mkern-2.5mu
      \in
      \mkern-2mu
      \mathrm{ran}(\hat{P})
      ,
      |\Psi\rangle
      \mkern-2.5mu
      \in
      \mkern-2mu
      \mathrm{ker}(\hat{P})
   \right)
   =
   \left\{
      \begingroup\SmallColSep
      \begin{array}{r l}
         1
         ,
         &
         \mkern15mu
         \text{if}
         \mkern4mu
         |\Psi\rangle
         \mkern-2.5mu
         \in
         \mkern-2mu
         \mathrm{ran}(\hat{P})
         \mkern6mu
         \text{is true}
         \\[5pt]
         0
         ,
         &
         \mkern15mu
         \text{if}
         \mkern4mu
         |\Psi\rangle
         \mkern-2.5mu
         \in
         \mkern-2mu
         \mathrm{ker}(\hat{P})
         \mkern6mu
         \text{is true}
         \\[5pt]
         0
         ,
         &
         \mkern15mu
         \text{if}
         \mkern4mu
         |\Psi\rangle
         \mkern-2.5mu
         \notin
         \mkern-2mu
         \mathrm{ran}(\hat{P})
         \sqcap
         |\Psi\rangle
         \mkern-2.5mu
         \notin
         \mkern-2mu
         \mathrm{ker}(\hat{P})
         \mkern6mu
         \text{is true}
      \end{array}
      \endgroup   
   \right.
   \;\;\;\;  ,
\end{equation}
\smallskip

\noindent or\smallskip

\begin{equation}  
   v
   \left(
      |\Psi\rangle
      \mkern-2.5mu
      \in
      \mkern-2mu
      \mathrm{ran}(\hat{P})
      ,
      |\Psi\rangle
      \mkern-2.5mu
      \in
      \mkern-2mu
      \mathrm{ker}(\hat{P})
   \right)
   =
   \left\{
      \begingroup\SmallColSep
      \begin{array}{r l}
         1
         ,
         &
         \mkern15mu
         \text{if}
         \mkern4mu
         |\Psi\rangle
         \mkern-2.5mu
         \in
         \mkern-2mu
         \mathrm{ran}(\hat{P})
         \mkern6mu
         \text{is true}
         \\[5pt]
         0
         ,
         &
         \mkern15mu
         \text{if}
         \mkern4mu
         |\Psi\rangle
         \mkern-2.5mu
         \in
         \mkern-2mu
         \mathrm{ker}(\hat{P})
         \mkern6mu
         \text{is true}
         \\[5pt]
         1
         ,
         &
         \mkern15mu
         \text{if}
         \mkern4mu
         |\Psi\rangle
         \mkern-2.5mu
         \notin
         \mkern-2mu
         \mathrm{ran}(\hat{P})
         \sqcap
         |\Psi\rangle
         \mkern-2.5mu
         \notin
         \mkern-2mu
         \mathrm{ker}(\hat{P})
         \mkern6mu
         \text{is true}
      \end{array}
      \endgroup   
   \right.
   \;\;\;\;  .
\end{equation}
\smallskip

\noindent Since\smallskip

\begin{equation}  
   |\Psi\rangle
   \mkern-2.5mu
   \in
   \mkern-2mu
   \mathrm{ker}(\hat{P})
   \sqcup
   \left(
      |\Psi\rangle
      \mkern-2.5mu
      \notin
      \mkern-2mu
      \mathrm{ran}(\hat{P})
      \sqcap
      |\Psi\rangle
      \mkern-2.5mu
      \notin
      \mkern-2mu
      \mathrm{ker}(\hat{P})
   \right)
   \mkern-2mu
   \iff
   \mkern-2mu
   |\Psi\rangle
   \mkern-2.5mu
   \notin
   \mkern-2mu
   \mathrm{ran}(\hat{P})
   \;\;\;\;  ,
\end{equation}
\\[-35pt]

\begin{equation}  
   |\Psi\rangle
   \mkern-2.5mu
   \in
   \mkern-2mu
   \mathrm{ran}(\hat{P})
   \sqcup
   \left(
      |\Psi\rangle
      \mkern-2.5mu
      \notin
      \mkern-2mu
      \mathrm{ran}(\hat{P})
      \sqcap
      |\Psi\rangle
      \mkern-2.5mu
      \notin
      \mkern-2mu
      \mathrm{ker}(\hat{P})
   \right)
   \mkern-2mu
   \iff
   \mkern-2mu
   |\Psi\rangle
   \mkern-2.5mu
   \notin
   \mkern-2mu
   \mathrm{ker}(\hat{P})
   \;\;\;\;  ,
\end{equation}
\smallskip

\noindent this would mean that the function $v$ could return the value of true or false depending only on the statement $|\Psi\rangle\mkern-2.5mu\in\mkern-2mu\mathrm{ran}(\hat{P})$ or the statement $|\Psi\rangle\mkern-2.5mu\in\mkern-2mu\mathrm{ker}(\hat{P})$ alone (not both of them). E.g.,\smallskip

\begin{equation} \label{QL} 
   {[\mkern-3.3mu[P]\mkern-3.3mu]}_v
   =
   v
   \left(
      |\Psi\rangle
      \mkern-2.5mu
      \in
      \mkern-2mu
      \mathrm{ran}(\hat{P})
   \right)
   =
   \left\{
      \begingroup\SmallColSep
      \begin{array}{r l}
         1
         ,
         &
         \mkern15mu
         \text{if}
         \mkern4mu
         |\Psi\rangle
         \mkern-2.5mu
         \in
         \mkern-2mu
         \mathrm{ran}(\hat{P})
         \mkern6mu
         \text{is true}
         \\[5pt]
         0
         ,
         &
         \mkern15mu
         \text{if}
         \mkern4mu
         |\Psi\rangle
         \mkern-2.5mu
         \notin
         \mkern-2mu
         \mathrm{ran}(\hat{P})
         \mkern6mu
         \text{is true}
      \end{array}
      \endgroup   
   \right.
   \;\;\;\;  .
\end{equation}
\smallskip

\noindent Seeing as

\begin{equation}  
   \mkern-2mu
   |\Psi\rangle
   \mkern-2.5mu
   \notin
   \mkern-2mu
   \mathrm{ran}(\hat{P})
   \iff
   U_{\mathbf{R}}
   =
   \varnothing
\;\;\;\;  ,
\end{equation}
\smallskip

\noindent one finds that in case Conjecture 3 held, a truth value of the experimental proposition $P$ in the state $|\Psi\rangle$ would be computed by the formula:\smallskip

\begin{equation} \label{QL1} 
   {[\mkern-3.3mu[P]\mkern-3.3mu]}_v
   =
   v
   \left(
      U_{\mathbf{R}}
      \neq
      \varnothing
   \right)
   =
   \left\{
      \begingroup\SmallColSep
      \begin{array}{r l}
         1
         ,
         &
         \mkern15mu
         \text{if}
         \mkern5mu
         U_{\mathbf{R}}
         \neq
         \varnothing
         \mkern6mu
         \text{is true}
         \\[5pt]
         0
         ,
         &
         \mkern15mu
         \text{if}
         \mkern5mu
         U_{\mathbf{R}}
         =
         \varnothing
         \mkern6mu
         \text{is true}
      \end{array}
      \endgroup   
   \right.
   \;\;\;\;  .
\end{equation}
\smallskip

\noindent As it has been demonstrated, to confirm the truth of the statements $U_{\mathbf{R}}\neq\varnothing$ and $U_{\mathbf{R}}=\varnothing$ requires $O(n)$ primitive operations for each. Therefore, according to (\ref{QL1}), the values of true and false of experimental quantum propositions would have no different cost of computation even without Conjecture 2.\\

\noindent However, unlike Conjecture 2, Conjecture 3 is \emph{unfalsifiable}. This means that even theoretically it would be impossible that Conjecture 3 could come into conflict with observation. More importantly, consequences of that conjecture cannot be demonstrated to be false by observation or experimentation.\\

\noindent To see this, take two experimental propositions $Q$ and $P$ associated with the projection operators $\hat{Q}$ and $\hat{P}$, which do not commute with each other, i.e., $\hat{Q}\hat{P}\neq\hat{P}\hat{Q}$. Suppose that the unit vector $|\Phi\rangle$ belongs to the range of the operator $\hat{Q}$; accordingly, the statement $|\Phi\rangle\mkern-2.5mu\in\mkern-2mu\mathrm{ran}(\hat{Q})$ is true and the statement $|\Phi\rangle\mkern-2.5mu\in\mkern-2mu\mathrm{ker}(\hat{Q})$ is false.\\

\noindent As it directly follows from the definition (\ref{CLS}), if $\hat{Q}\hat{P}|\Phi\rangle\neq\hat{P}\hat{Q}|\Phi\rangle$, then $|\Phi\rangle\mkern-2.5mu\notin\mkern-2mu\mathrm{ran}(\hat{P})$ and $|\Phi\rangle\mkern-2.5mu\notin\mkern-2mu\mathrm{ker}(\hat{P})$.\\

\noindent Along the lines of \cite{Birkhoff}, assume that $Q \sqcap P$ is represented by the lattice-theoretic meet\smallskip

\begin{equation}  
   \mathrm{ran}(\hat{Q})
   \wedge
   \mathrm{ran}(\hat{P})
   \equiv
   \mathrm{ran}(\hat{Q})
   \cap
   \mathrm{ran}(\hat{P})
   =
   \left\{
      |\psi\rangle
      \in
      \mathcal{H}
      \textnormal{:}
      \mkern10mu
      \mathfrak{F}
      \left(
         |\psi\rangle
      \right)
   \right\}
   \;\;\;\;  ,
\end{equation}
\smallskip

\noindent where the predicate $\mathfrak{F}(|\psi\rangle)$ is the conjunction of two valuations, namely,\smallskip

\begin{equation}  
   \mathfrak{F}
   \left(
      |\psi\rangle
   \right)
   =
   v
   \mkern-2.5mu
   \left(
      |\psi\rangle
      \mkern-2.5mu
      \in
      \mkern-2mu
      \mathrm{ran}(\hat{Q})
   \right)
   \sqcap
   v
   \mkern-2.5mu
   \left(
      |\psi\rangle
      \mkern-2.5mu
      \in
      \mkern-2mu
      \mathrm{ran}(\hat{P})
   \right)
   \;\;\;\;  .
\end{equation}
\smallskip

\noindent Because $|\Phi\rangle\mkern-2.5mu\notin\mkern-2mu\mathrm{ran}(\hat{P})$ is true, the value of $v$ at $|\Phi\rangle\mkern-2.5mu\in\mkern-2mu\mathrm{ran}(\hat{P})$ is false, as stated by (\ref{QL}). Consequently, the non-zero vector $|\Phi\rangle$ cannot be an element of the meet $\mathrm{ran}(\hat{Q})\cap\mathrm{ran}(\hat{P})$, which indicates that this meet coincides with the zero subspace $\{0\}$ or, equivalently, $\mathrm{ran}(\hat{0})$. Thus, for any non-zero vector $|\Psi\rangle$, the following must hold:\smallskip

\begin{equation} \label{CONJ1} 
   {[\mkern-3.3mu[Q \sqcap P]\mkern-3.3mu]}_v
   =
   v
   \left(
      |\Psi\rangle
      \mkern-2.5mu
      \in
      \mkern-2mu
      \mathrm{ran}(\hat{0})
   \right)
   =
   0
   \;\;\;\;  .
\end{equation}
\smallskip

\noindent Likewise, on condition that $Q \sqcap \neg{P}$ is represented by\smallskip

\begin{equation}  
   \mathrm{ran}(\hat{Q})
   \wedge
   \mathrm{ran}(\hat{1}-\hat{P})
   \equiv
   \mathrm{ran}(\hat{Q})
   \cap
   \mathrm{ker}(\hat{P})
   =
   \left\{
      |\psi\rangle
      \in
      \mathcal{H}
      \textnormal{:}
      \mkern10mu
      \mathfrak{F}^{\mkern2mu\prime}
      \left(
         |\psi\rangle
      \right)
   \right\}
   \;\;\;\;  ,
\end{equation}
\smallskip

\noindent where

\begin{equation}  
   \mathfrak{F}^{\mkern2mu\prime}
   \left(
      |\psi\rangle
   \right)
   =
   v
   \left(
      |\psi\rangle
      \mkern-2.5mu
      \in
      \mkern-2mu
      \mathrm{ran}(\hat{Q})
   \right)
   \sqcap
   v
   \left(
      |\psi\rangle
      \mkern-2.5mu
      \in
      \mkern-2mu
      \mathrm{ker}(\hat{P})
   \right)
   \;\;\;\;  ,
\end{equation}
\smallskip

\noindent one finds that in any physically meaningful state $|\Psi\rangle$ the following must hold:\smallskip

\begin{equation} \label{CONJ2} 
   {[\mkern-3.3mu[Q \sqcap \neg{P}]\mkern-3.3mu]}_v
   =
   v
   \left(
      |\Psi\rangle
      \mkern-2.5mu
      \in
      \mkern-2mu
      \mathrm{ran}(\hat{0})
   \right)
   =
   0
   \;\;\;\;  .
\end{equation}
\smallskip

\noindent But then, because $\hat{Q}$ and $\hat{P}$ do not commute, the experimental propositions $Q$ and $P$, as well as $Q$ and $\neg{P}$, cannot be verified simultaneously, and, as a result, the equalities (\ref{CONJ1}) and (\ref{CONJ2})  cannot be testable or refutable by experiment.\\

\noindent Providing – again, in line with \cite{Birkhoff} – that the disjunction $P \sqcup \neg{P}$ is represented by the lattice-theoretic join\smallskip

\begin{equation}  
   \mathrm{ran}(\hat{P})
   \vee
   \mathrm{ker}(\hat{P})
   =
   \mathrm{ran}(\hat{P})
   \oplus
   \mathrm{ker}(\hat{P})
   =
   \mathrm{ran}(\hat{1})
   \;\;\;\;  ,
\end{equation}
\smallskip

\noindent the failure of the distributive law of propositional logic immediately follows from Conjecture 3. Really, since\smallskip

\begin{equation}  
   \mathrm{ran}(\hat{Q})
   \cap
   \mathrm{ran}(\hat{1})
   =
   \mathrm{ran}(\hat{Q})
   \;\;\;\;  ,
\end{equation}
\smallskip

\noindent one gets $Q \sqcap (P \sqcup \neg{P}) = Q$, which is true in the state $|\Phi\rangle$, whereas $Q \sqcap P$ or $Q \sqcap \neg{P}$ is always false.\\

\noindent Given that the breakdown of the distributive law is one of most notable properties of the traditional quantum logic (QL) of Birkhoff and von Neumann \cite{Birkhoff, Piron}, Conjecture 3 can be referred to as \emph{QL conjecture}.\\

\noindent Even if one does not see falsifiability as an indispensable principle for evaluating every logic used to handle the statements and propositions of the theories employed to describe the physical world, one cannot ignore theoretical objections to QL.\\

\noindent Particularly, QL is too radical in giving up distributivity: By doing so, QL causes the interpretation of the logical operations $\sqcap$ and $\sqcup$ as conjunction and disjunction to become problematic \cite{Griffiths}. On the other hand, QL is not radical enough in insisting that the valuation $v$ can only be a total Boolean function: Upon doing so, QL falls victim to Schrödinger's cat and the like \cite{Caspers}.\\

\noindent Therefore, unless the QL conjecture (unrefutable by experiment) is assumed, the same cost of computation for each truth value of an experimental quantum proposition must entail the existence of based on quantum mechanics parallel computing whose efficiency can be greater than 1.\\

\bibliographystyle{References}
\bibliography{Valuation}

\end{document}